\documentclass[a4paper,aps,pre,superscriptaddress,floatfix,nofootinbib,longbibliography,notitlepage,onecolumn]{revtex4-1}
\usepackage{bbm}
\usepackage{bbold}
\usepackage{bbm}
\usepackage{graphicx}
\usepackage{latexsym,amsmath,verbatim,amssymb,txfonts}
\usepackage{lipsum}
\usepackage{color}
\usepackage{rotating}
\usepackage{verbatim}
\usepackage{multirow}
\usepackage[english]{babel}
\usepackage{comment}
\usepackage{bm}
\usepackage{mathtools}
\usepackage{amsmath}
\usepackage{amsfonts}
\usepackage{amssymb}

\begin{document}
\title{Time delay embeddings to characterize the timbre of musical instruments using Topological Data Analysis: a study on synthetic and real data}

\author{Gakusei Sato}
\email{gssato.ac@gmail.com}
\affiliation{Department of Systems and Control Engineering, Institute of Science Tokyo (former Tokyo Tech), Tokyo, Japan}

\author{Hiroya Nakao}
\affiliation{Department of Systems and Control Engineering, Institute of Science Tokyo (former Tokyo Tech), Tokyo, Japan}
\affiliation{International Research Frontiers Initiative, Institute of Science Tokyo (former Tokyo Tech), Kanagawa, Japan}

\author{Riccardo Muolo}
\affiliation{Department of Systems and Control Engineering, Institute of Science Tokyo (former Tokyo Tech), Tokyo, Japan}

\date{\today}

\begin{abstract}

Timbre allows us to distinguish between sounds even when they share the same pitch and loudness, playing an important role in music, instrument recognition, and speech. Traditional approaches, such as frequency analysis or machine learning, often overlook subtle characteristics of sound. Topological Data Analysis (TDA) can capture complex patterns, but its application to timbre has been limited, partly because it is unclear how to represent sound effectively for TDA. In this study, we investigate how different time delay embeddings affect TDA results. Using both synthetic and real audio signals, we identify time delays that enhance the detection of harmonic structures. Our findings show that specific delays, related to fractions of the fundamental period, allow TDA to reveal key harmonic features and distinguish between integer and non-integer harmonics. The method is effective for synthetic and real musical instrument sounds and opens the way for future works, which could extend it to more complex sounds using higher-dimensional embeddings and additional persistence statistics.


\end{abstract}

\maketitle

\section{Introduction}

Timbre, together with loudness and pitch, is a fundamental aspect of sound that allows us to distinguish between different sources even when they have the same loudness and pitch~\cite{JIS2000}. It is an important factor in applications such as music information retrieval~\cite{Tzanetakis2002} and separating speakers in audio recordings~\cite{Ohi2020}. A key characteristic of timbre is its harmonic content. This includes integer harmonics, which are exact multiples of the fundamental frequency and result from the physical mechanisms of sound production, as well as non-integer harmonics, which can arise from effects like how a string is plucked or variations in airflow. The combination of these harmonics contributes to the richness and complexity of the timbre.

Traditional timbre analysis usually uses frequency-based measures such as sharpness~\cite{Bismarck1974} and spectral flatness~\cite{Johnston1988}. Machine learning approaches have been used to improve feature extraction for tasks like sound classification and synthesis~\cite{Cosi1998,Pons2017}. Despite these methods, it remains difficult to fully capture the perceptual richness of timbre, since its harmonic characteristics result from complex interactions in the sound production process. Adavanced data analysis tools are needed to obrain more accurate results.

For example, network science~\cite{latora_nicosia_russo_2017,newmanbook} provides a powerful tool to analyze data~\cite{ortega2018graph}, with some limitations. The recent introduction of higher-order networks~\cite{battiston2020networks,bick2023higher,muolo2024turing} extends the classical network approach by incorporating interactions beyond pairwise links, which capture many-body relationships present in many real-world systems. The effects of higher-order interactions have been studied in the context of dynamics, e.g., phase models~\cite{tanaka2011multistable,skardal2020higher,leon2024}, synchronization~\cite{gambuzza2021stability,gallo2022synchronization,della2023emergence}, random walk~\cite{carletti2020random,schaub2020random}, or pattern formation~\cite{carletti2020dynamical,muologallo}, to name a few. But their applications go well beyond dynamics. In fact, the mathematical tools of higher-order structures are extremely useful in data analysis, and have been applied to various domains, from neural networks~\cite{zhou2007learning,roddenberry2019hodgenet,hajij2020cell,ebli2020simplicial} to signal processing~\cite{schaub2021signal,sardellitti2022topological,calmon2023dirac,sardellitti2024topological,wang2025dirac}, demonstrating their power and ductility to reveal hidden structures in high dimensional data. 

One of the most promising applications of higher-order networks is in Topological Data Analysis (TDA)~\cite{Edelsbrunner2002,Patania2017,wei2023characterizing,Carlsson2009}, which sets a rigorous mathematical framework to extract the shape of data. TDA captures topological features such as connected components, cycles (i.e., $1$-dimensional holes), and higher-dimensional holes, identifying the structural properties. This makes TDA particularly suited for analyzing high-dimensional datasets, where classical methods may fail to reveal meaningful features. Recent works have also refined the use of persistent homology by introducing the persistent Dirac-Bianconi operator~\cite{wee2023persistent} for representing and analyzing data with intricate geometric and topological patterns.

TDA has also proven ductile and powerful in the study of time series data, especially when integrated with machine learning techniques~\cite{ElYaagoubi2023,Seversky2016,Perea2015}. For example, TDA has been applied to the analysis of musical instrument sounds, showing that topological features could improve classification accuracy compared to traditional frequency-based descriptors~\cite{Sanderson2017}. TDA was also applied to characterize musical sounds, showing how topological invariants (e.g., persisten homology) can capture timbral and harmonic structures that are otherwise difficult to quantify~\cite{wei2023characterizing}. Such studies put to the fore the potential of TDA in uncovering structural patterns in data which are not easily captured by classical approaches. In general, topological approaches find many applications. In neuroscience, for example, the topology of neural activity and connectivity patterns has been shown to encode functionally relevant information~\cite{Bassett,petri2014homological,expert2019topological}. Similarly, in other fields such as genomics, social systems, and materials science, the structural insights provided by topological methods are becoming more and more relevant~\cite{Patania2017,patania2017shape}. 

In this work, we use TDA to look at harmonic structures in musical signals to characterize timbre. We systematically investigate the effects of embedding parameters to the topological features extracted from time series data, providing a thorough evaluation of their impact on capturing timbral characteristics. Our numerical experiments, conducted on both synthetic and real-world musical datasets, show that a fine tuning of these parameters can significantly enhance the sensitivity of TDA to harmonic structures. This approach offers an intriguing perspective on harmonic analysis and provides new tools for exploring the topology of sound data.

The paper is structured as follows. In the next Section, we will give an introduction of the framework of higher-order networks and the main mathematical tools. Then, in Sec. \ref{sec:Methods}, we will introduce Topological Data Analysis (TDA) and the embedding methods. All numerical results are shown and discussed in Sec. \ref{sec:Results}, right before concluding and discussing future improvement and perspectives of the framework hereby provided.

\section{Introduction to the topology of higher-order networks}\label{sec:introTDA}

\begin{figure}[h!]
    \centering
    \includegraphics[scale=0.5]{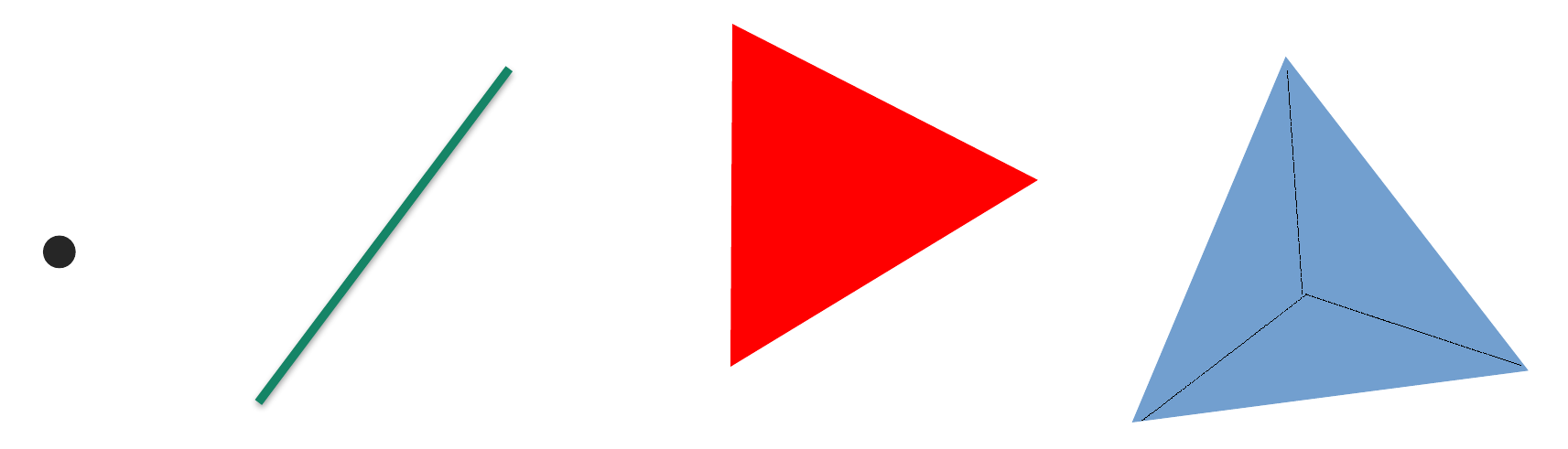}
    \caption{\textbf{Examples of $n$-simplices.} From left to right: a $0$-simplex, a $1$-simplex, a $2$-simplex, and a $3$-simplex. Note that in a simplicial complex one cannot find an isolated $n$-simplex (for $n\geq 1$), as the presence of an $n$-simplex implies the existence of all the $(n-1)$-simplices.}    
    \label{fig_B}
\end{figure}

The classic network framework consists on pairwise interactions between nodes, encoded by links. This approach has proven ductile and powerful and has greatly contributed in the advances of complex systems~\cite{latora_nicosia_russo_2017,newmanbook}. However, interactions in many real-world systems are \emph{higher-order}, i.e., involving groups of nodes simultaneously~\cite{battiston2020networks,bick2023higher,muolo2024turing}. Mathematically, such group interactions are encoded by \textit{hypergraphs} and \textit{simplicial complexes}~\cite{natphys}. In this work, we focus on the latter, due to their mathematical framework based on algebraic topology~\cite{Lim2020,bianconi2021higher,millan2025topology}.
A \emph{simplicial complex} $\mathcal{K}$ is a set of simplices of various dimensions, i.e., $0$-simplices, corresponding to \emph{nodes}, $1$-simplices, corresponding to \emph{edges}, $2$-simplices, corresponding to \emph{triangles}, $3$-simplices, corresponding to \emph{tetrahedra}, and so on. Examples of simplices are depicted in Fig. \ref{fig_B}.

A simplex is formally defined as an ordered set of $0$-simplices $\sigma = [v_0, v_1, \dots, v_k ]$, where $k$ denotes its dimension. Observe that a simplex is always defined by means of the $0$-simplices, even though the latter are not a part of the simplex itself. A simplicial complex satisfies the property of closure: if $\sigma \in \mathcal{K}$, then all faces of $\sigma$ (all subsets of its vertices) also belong to $\mathcal{K}$. For example, the presence of a triangle ($2$-simplex) implies that its three edges ($1$-simplices) and three vertices ($0$-simplices) are also part of the simplicial complex. 

The structure of a simplicial complex can be encoded via boundary operators $\mathbf{B}_k$, which map each $k$-simplex to the formal sum of its $(k-1)$-dimensional faces:
\begin{equation}
\mathbf{B}_k : C_k(\mathcal{K}) \to C_{k-1}(\mathcal{K}), \quad 
\mathbf{B}_k \{v_0, \dots, v_k\} = \sum_{i=0}^{k} (-1)^i \{v_0, \dots, \hat{v}_i, \dots, v_k\},
\end{equation}
where $C_k(\mathcal{K})$ denotes the space of $k$-chains and $\hat{v}_i$ indicates omission of the $0$-simplex $v_i$. 
In more practical terms, given $N_k$ $k$-simplices and $N_{k-1}$ $(k-1)$-simplices, $\mathbf{B}_k$ is a $N_{k-1} \times N_k$ matrix defined as
\begin{equation} \label{matrixBoundary}
    \mathbf{B}_k(i,j)=
    \begin{cases}
        1 & \text{if the $i$-th $(k-1)$-simplex is a face of the $j$-th $k$-simplex with same orientation,}\\
        -1 & \text{if the $i$-th $(k-1)$-simplex is a face of the $j$-th $k$-simplex with opposite orientation,}\\
        0 & \text{if the $i$-th $(k-1)$-simplex is not a face of the $j$-th $k$-simplex.}
    \end{cases}
\end{equation}

For $k=1$, one obtains the well-known incidence matrix for networks~\cite{newmanbook}. Note that $\mathbf{B}_{k-1} \circ \mathbf{B}_k = 0$, meaning that the boundary of a boundary is empty.

\begin{figure}[h!]
    \centering
    \includegraphics[scale=0.4]{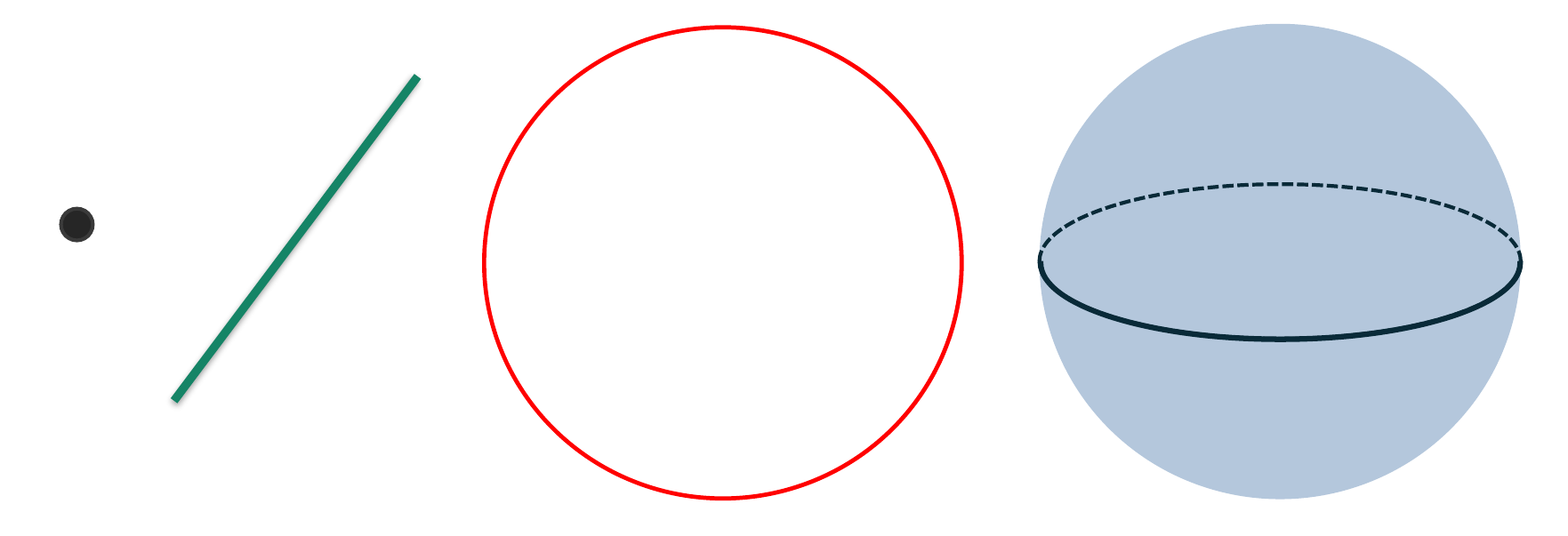}
    \caption{\textbf{Examples of topological structures and their Betti numbers.} From left to right: a dot ($0$-dimensional structure), whose only non-zero Betti number is $\beta_0=1$ (i.e., one connected component); a line ($1$-dimensional), with $\beta_0=1$ (i.e., one connected component); a circle ($2$-dimensional, with $\beta_0=1$ and $\beta_1=1$ (i.e., one connected component and one hole); an empty sphere ($3$-dimensional, with $\beta_0=1$, $\beta_1=0$, and $\beta_2=1$ (i.e., one connected component, no holes, and one empty volume).}    
    \label{fig_A}
\end{figure}

\begin{figure}[h!]
    \centering
    \includegraphics[scale=0.5]{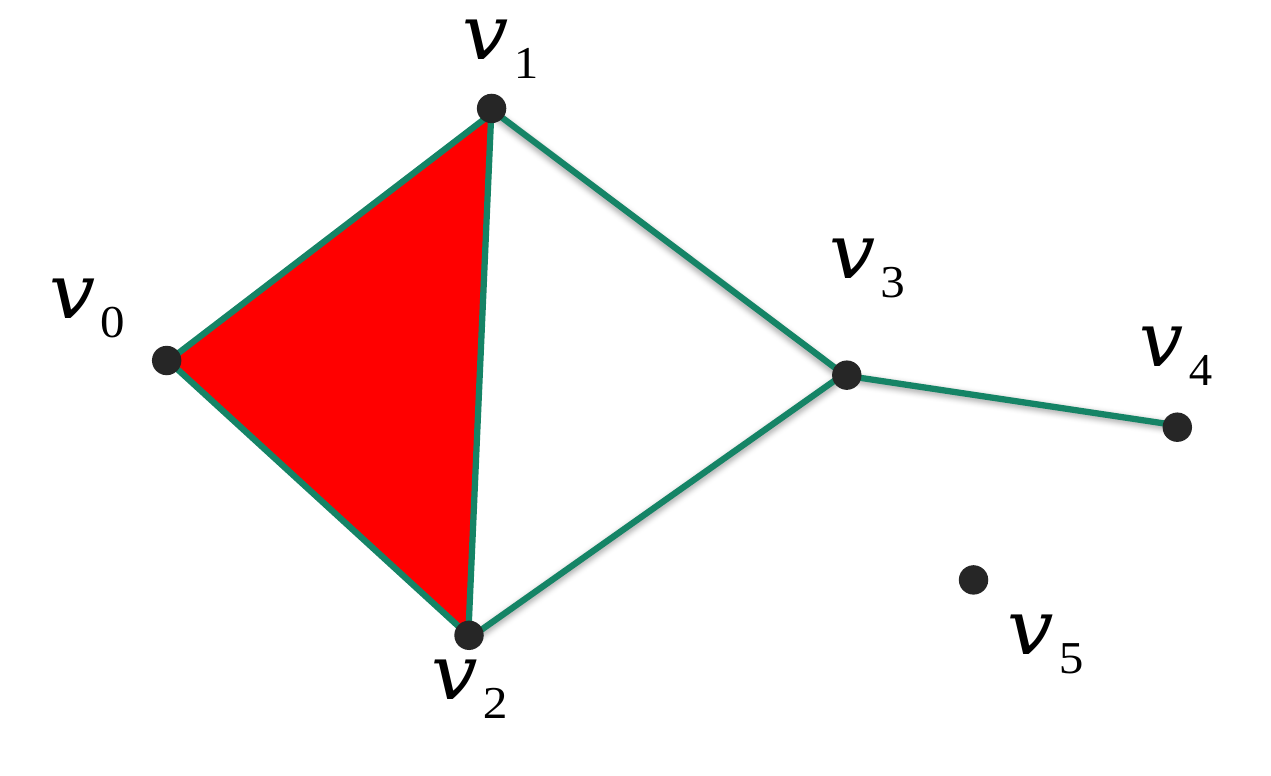}
    \caption{\textbf{Topology of a simplicial complex.} Example of a simplicial complex made of $6$ nodes ($0$-simplices) $[v_0],[v_1],[v_2],[v_3],[v_4],[v_5]$, $6$ links ($1$-simplices) $[v_0,v_1],[v_0,v_2],[v_1,v_2],[v_1,v_3],[v_2,v_3],[v_3,v_4]$, and $1$ triangle ($2$-simplex) $[v_0,v_1,v_2]$. There are $2$ connected components, so $\beta_0=2$, and one hole, so that $\beta_1=1$. Note that this simplicial complex is topologically equivalent to another simplicial complex obtained with only $[v_1],[v_2],[v_3],[v_5]$ as $0$-simplices, and $[v_1,v_2],[v_1,v_3],[v_2,v_3]$ as $1$-simplices, as the latter would have the same Betti numbers $\beta_0$ and $\beta_1$.}    
    \label{fig_C}
\end{figure}

The topological features of the complex are captured by the homology groups
\begin{equation}
H_k(\mathcal{K}) = \ker \mathbf{B}_k \,/\, \operatorname{im} \mathbf{B}_{k+1},
\end{equation}
which identify $k$-dimensional cycles not bounding a $(k+1)$-simplex. The ranks of these groups, called Betti numbers $\beta_k = \mathrm{rank}\, H_k(\mathcal{K})$, quantify the number of $k$-dimensional topological features. For instance, $\beta_0$ indicates the number of connected components, $\beta_1$ the number of cycles (i.e., holes), $\beta_2$ counts empty volumes, and so on. Examples of structures with their respective Betti numbers are shown in Fig. \ref{fig_A}, while an example of simplicial complex is shown in Fig. \ref{fig_C}. In simplicial complexes, it is important to note the difference between a cycle and a boundary. Looking at Fig. \ref{fig_C}, we might observe that there are two cycles, namely $[v_0,v_1],[v_0,v_2],[v_1,v_2]$ and $[v_1,v_2],[v_1,v_3],[v_2,v_3]$. However, while the latter is a cycles, the former is the boundary of the $2$-simplex $[v_0,v_1,v_2]$, hence it does not contribute to the Betti number $\beta_1$ because it is not a hole (roughly speaking, the "space" is filled by the $2$-simplex).

Betti numbers are topological invariants that allow us to determine whether two structures are topologically equivalent, i.e., homeomorphic, meaning that they can be continuously deformed to obtain the other\footnote{The most famous example is the homeomorfism between a mug and a doughnut, both having $\beta_0=1$ and $\beta_1=1$.}. For simplicial complexes, Betti numbers can be computed from linear algebra~\cite{Lim2020}. 
To do that, we need to define the coboundary operator. This would require the introduction of other mathematical tools, which we do not need for our analysis. Hence, we will keep the mathematical jargon to the minimum. The interested reader my consult a textbook~\cite{bianconi2021higher} and a review~\cite{Lim2020}, where such concepts are introduced formally. Let us introduce the transposed of the boundary operator $\mathbf{B}_k^\top$, which is called coboundary operator. Using the boundary and coboundary operators, we can define the Hodge Laplacian, models diffusion among $k$-simplices, as
\begin{equation}
\mathbf{L}_k = \mathbf{B}_{k+1} \mathbf{B}_{k+1}^\top + \mathbf{B}_k^\top \mathbf{B}_k.
\end{equation}
Note that $L_0$ is the well-known graph Laplacian~\cite{newmanbook}. The Betti number $\beta_k$ is the dimension of the kernel of $L_k$, which can be computed quite easily\footnote{For those familiar with network science, it is known that the Laplacian $L_0$ has as many zero eigenvalues as there are connected components in the networks~\cite{newmanbook}. This is exactly the Betti number $\beta_0$.}. 

\section{Methods}\label{sec:Methods}

\subsection{Topological Data Analysis (TDA)}

\begin{figure}[h!]
    \centering
    \includegraphics[scale=0.5]{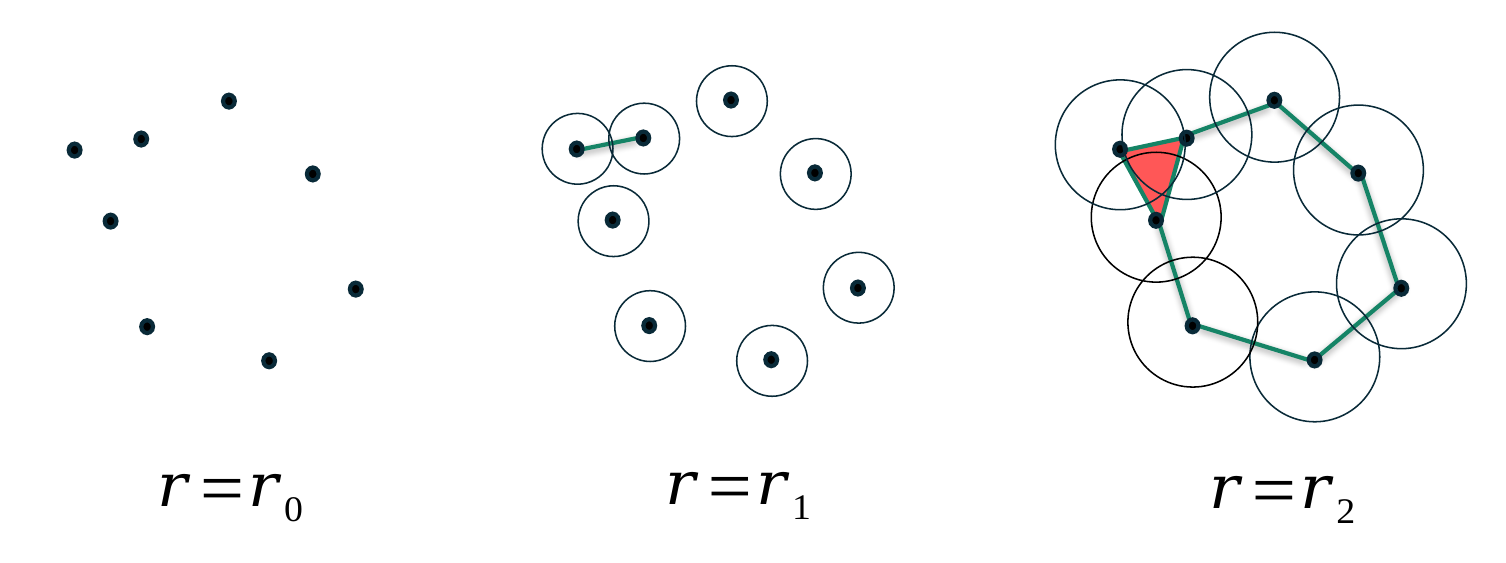}
    \caption{\textbf{Example of how a simplicial complex is extracted from data.} We start with a radius $r_0=0$, so all the points ($0$-simplices) are separated (left panel) and $\beta_0$ is equal to the number of points. Then we increase the radius making the circles around the point larger. Once two circles intersect, the two corresponding $0$-simplices are connected via a $1$-simplex (middle panel), and $\beta_0$ decreases. If three circles intersect simultaneously, then the corresponding $0$-simplices are connected via a $2$-simplex (right panel). In the latter case, we have that $\beta_0=1$ and $\beta_1=1$. In general, when $(n+1)$ circles intersect simultaneously, the corresponding $0$-simplices are connected via an $n$-simplex. The results of such analysis can be visualized through the \emph{persistence diagram} (see Fig. \ref{fig_E}).}    
    \label{fig_D}
\end{figure}

The concepts of the above Section can be exploited via Topological Data Analysis (TDA), which is a computational framework that applies algebraic topology to the study of high dimensional datasets~\cite{Patania2017}. Unlike traditional statistical or geometric methods, which often rely on linear projections or kernel-based techniques, TDA focuses on the intrinsic shape of data, i.e., their topological features. By leveraging topological invariants (i.e., Betti numbers), TDA provides a robust, multi-scale characterization of data structure that is robust to noise and does not require explicit parametric assumptions about the underlying distribution~\cite{Patania2017}.
A central tool in TDA is persistent homology, a method that computes topological invariants such as connected components, cycles, empty volumes, and higher-dimensional voids. Given a point cloud, TDA constructs a filtered simplicial complex, i.e., a representation of the data topology in form of a simplicial complex, whose structure was described in the previous section. A commonly used filtration method is the Vietoris–Rips complex, where a simplex $\sigma$ is included if all distances between its $0$-simplices are below a threshold $r$, representing the radius of the circle drawn around each $0$-simplex. As $r$ increases, the homology of the complex varies, capturing the formation of topological structure and analyzing their persistence via the corresponding Betti number. An example of this process for data points in a $2$-dimensional domain is shown in Fig. \ref{fig_D}. Note that throughout this work, we will only consider analogous settings.

\begin{figure}[h!]
    \centering
    \includegraphics[scale=0.5]{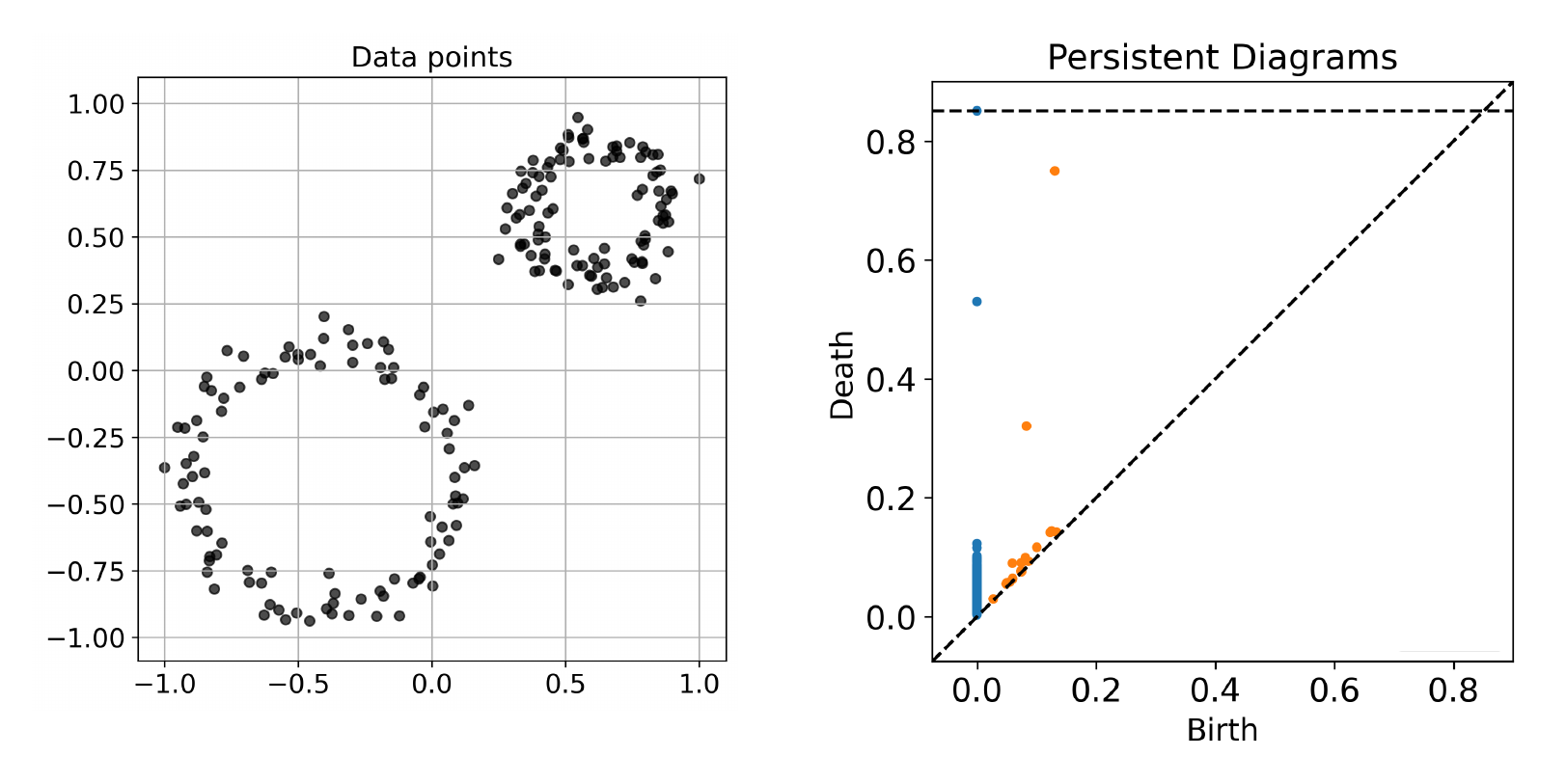}
    \caption{\textbf{$2$-dimensional data and persistence diagram.} On the left, example of data points in a $2$-dimensional domain. On the right, example of a persistence diagram, which summarizes the topological features of the dataset on the left panel across scales. Each point represents a topological feature. The $x$ axis indicates the radius at which the feature appears (“birth”), and the $y$ axis the radius at which it disappears (“death”). Topological features that persist with respect to the radius, i.e., far from the diagonal, correspond to meaningful structures in the data, while those close to the diagonal are usually caused by noise in the data.}    
    \label{fig_E}
\end{figure}

The birth and death of these features are recorded in a persistence diagram or barcode, providing a compact, multi-scale summary of the topological structure of the data. An example is shown in Fig. \ref{fig_E}. Note that, for data points in a $2$-dimensional domain, the filtered simplicial complex will have only $2$ non-zero Betti numbers, namely, $\beta_0$ and $\beta_1$.

One of the strengths of TDA lies in its ability to analyze high dimensional and unstructured data, while remaining robust to perturbations and noise. Persistent homology ensures that small perturbations in the input data induce only minor changes in the persistence diagram~\cite{CohenSteiner2005}. This property is particularly relevant in domains where data is intrinsically noisy, such as functional brain networks, molecular dynamics, and sensor arrays, to name a few~\cite{Patania2017}. For these reasons, TDA finds applications in many domains. For example, in neuroscience, it has been used to model the topology of functional connectivity networks, revealing higher-order structures beyond pairwise correlations~\cite{Giusti2016}. In genomics, persistent homology aids in analyzing DNA folding and chromatin interactions~\cite{Yao2009}. In image processing, TDA provides topological descriptors for feature extraction in shape recognition tasks~\cite{Carlsson2009}. Additionally, in time-series analysis, TDA has been applied to dynamical systems by leveraging delay embeddings, where the theoretical results that have been developed ensure that the topological invariants can be recovered in an appropriately chosen phase space~\cite{Perea2015}.
However, one of the main challenges of using TDA to analyze time series is the choice of the space where the data is embedded and analyzed. In fact, such choice is arbitrary and there is no general criterion for that. This motivates our study, whose main focus is to find suitable ways in which time series from musical instruments can be embedded in a $2$-dimensional space in order to maximize the impact of TDA on the analysis.

\subsection{Extracting the timbre}

In this study, we propose a method for extracting timbral features from audio signals using TDA. Since TDA generally assumes high dimensional point cloud data as input, it is necessary to reconstruct the one-dimensional time series audio signal into a high dimensional space. To achieve this, we employ time delay embedding~\cite{takens2006detecting} to transform the audio signal into a suitable format for TDA.

\begin{align}
\mathbf{X}_d(x_t; \tau) = \left(x_t, x_{t+\tau}, x_{t+2\tau}, \dots, x_{t+(d-1)\tau} \right),
\label{eq:tde}
\end{align}

\noindent where $x_t$ represents the time-series data to be analyzed, $\mathbf{X}_d$ is the embedding vector of dimension $d$, and $\tau$ is the time delay. This method reconstructs the time-series data as vectors corresponding to the embedding dimension $d$ while sequentially shifting them by a time delay $\tau$. Consequently, the audio data is embedded into a high-dimensional point cloud, which serves as the expected input shape for TDA. TDA is then applied to extract topological features, represented as a persistent diagram. Following previous studies~\cite{Sanderson2017}, the audio data is embedded in a two-dimensional space, and we focus on features related to first-order homology. In principle, higher dimensions could be considered, but at a high computational cost.

The objective of this study is to conduct a qualitative analysis of timbral structures using TDA. The persistent diagram is a set of points, and its correspondence to timbre-related features remains unclear. Timbre is generally characterized as the set of distinctive features differentiating an audio signal from a sine wave with the same fundamental frequency. Therefore, we quantify the difference in persistent diagrams between the analyzed audio signal and a sine wave of the same fundamental frequency using the Wasserstein distance. To do this, we can define the topological feature of timbre $m$ as follows:

\begin{align}
m = W(D(X_2(s)), D(X_2(s_0))),
\label{eq:pm}
\end{align}

\noindent where $s$ represents the analyzed audio data, $s_0$ denotes a sine wave with the same fundamental frequency as $s$, $X_2$ corresponds to the two-dimentional time delay embedding, $D$ represents the persistent diagram associated with first-order homology extracted via TDA, and $W$ is the Wasserstein distance as follows~\cite{Rubner2000,Mileyko2011}:

\begin{align}
W(D_1, D_2) = \inf_{\gamma \in \Gamma(D_1, D_2)} \sum_{(x, y) \in \gamma} |x - y|,
\label{eq:wd}
\end{align}
where $\Gamma(D_1, D_2)$ denotes the set of all bijections between the points of $D_1$ and $D_2$. This metric quantifies the difference between two persistent diagrams and enables the evaluation of topological features associated with timbre.

In time delay embedding, the delay parameter $\tau$ is a critical factor and must be chosen appropriately depending on the task~\cite{small2005applied, fraser1986independent}. However, in the context of applying TDA to timbral analysis, there is no established method for determining the optimal time delay. To address this, we introduce a synthetic audio signal $s(t; a)$ with an adjustable harmonic strength $a \in [0, 1]$. We then evaluate how the topological feature $m$ varies with different delay values $\tau$ as follows:

\begin{align}
m(a, \tau) = W(D(X_2(s(t; 0); \tau)), D(X_2(s(t; a); \tau))).
\label{eq:pm2}
\end{align}

This measure captures, via TDA, the structural change in topology induced by the presence of harmonic components. The implementation of this method is available at \url{https://github.com/gssato/TopologicalHarmonicExtraction.git}.

\section{Numerical Results}\label{sec:Results}

\subsection{Synthetic data}

In this Section, we validate the effectiveness of the proposed method through numerical experiments using synthesized audio signals. First, we present the results of applying time delay embedding and TDA to several synthesized signals $s(t)$. 

\begin{figure} [h!] 
    \centering
    \includegraphics[scale=0.5]{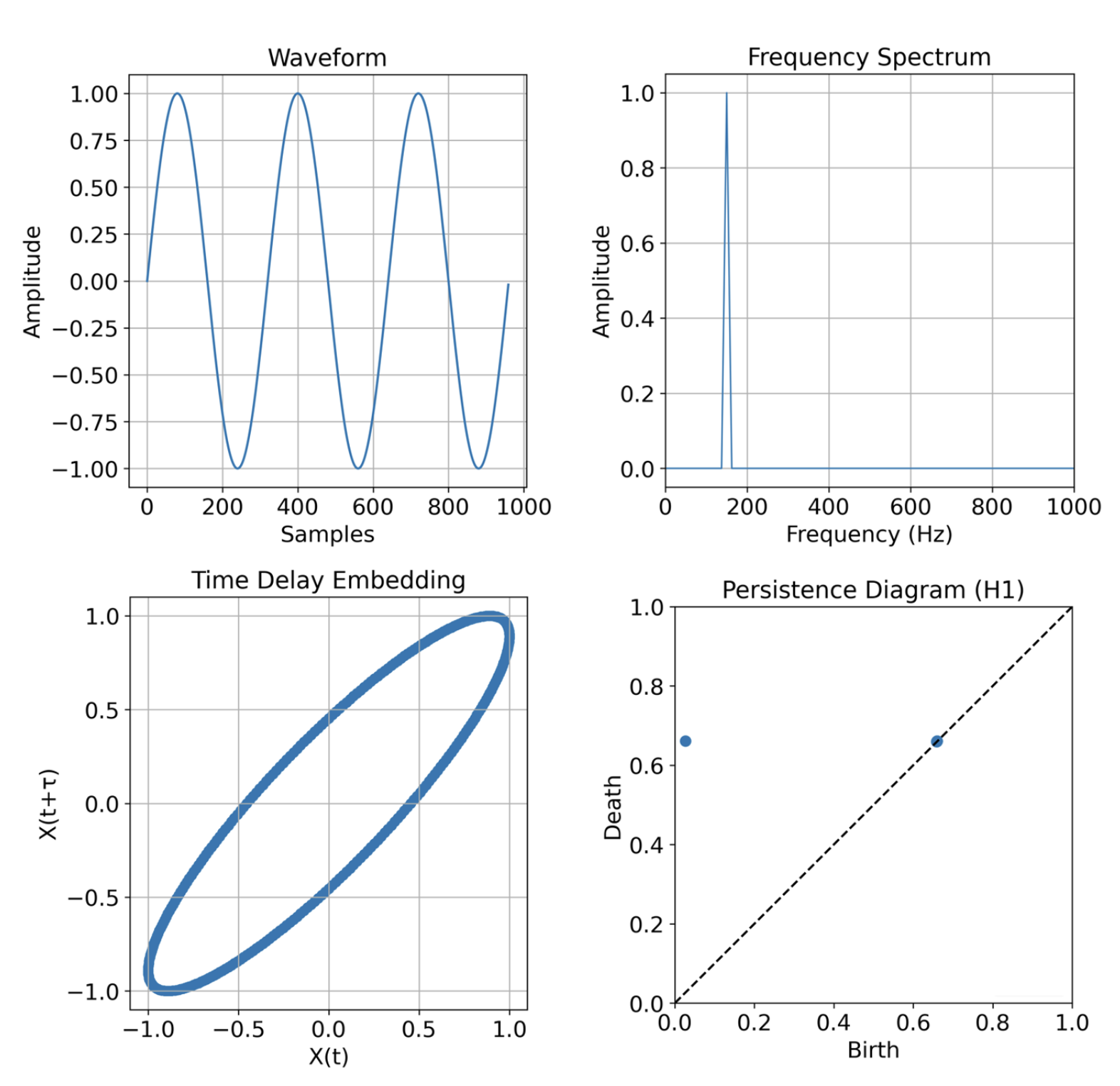}
            \caption{\textbf{TDA of a harmonic signal $s(t)=\sin(2\pi f_0 t)$.} The upper-left panel shows the waveform, the upper-right panel shows the frequency spectrum, the lower-left panel shows the time delay embedding, and the lower-right panel shows the persistence diagram. The embedding forms a regular circular trajectory, and the persistence diagram captures this structure as a point located far from the diagonal.}
    \label{fig:s1_1}
\end{figure}

\begin{figure} [h!] 
    \centering
    \includegraphics[scale=0.5]{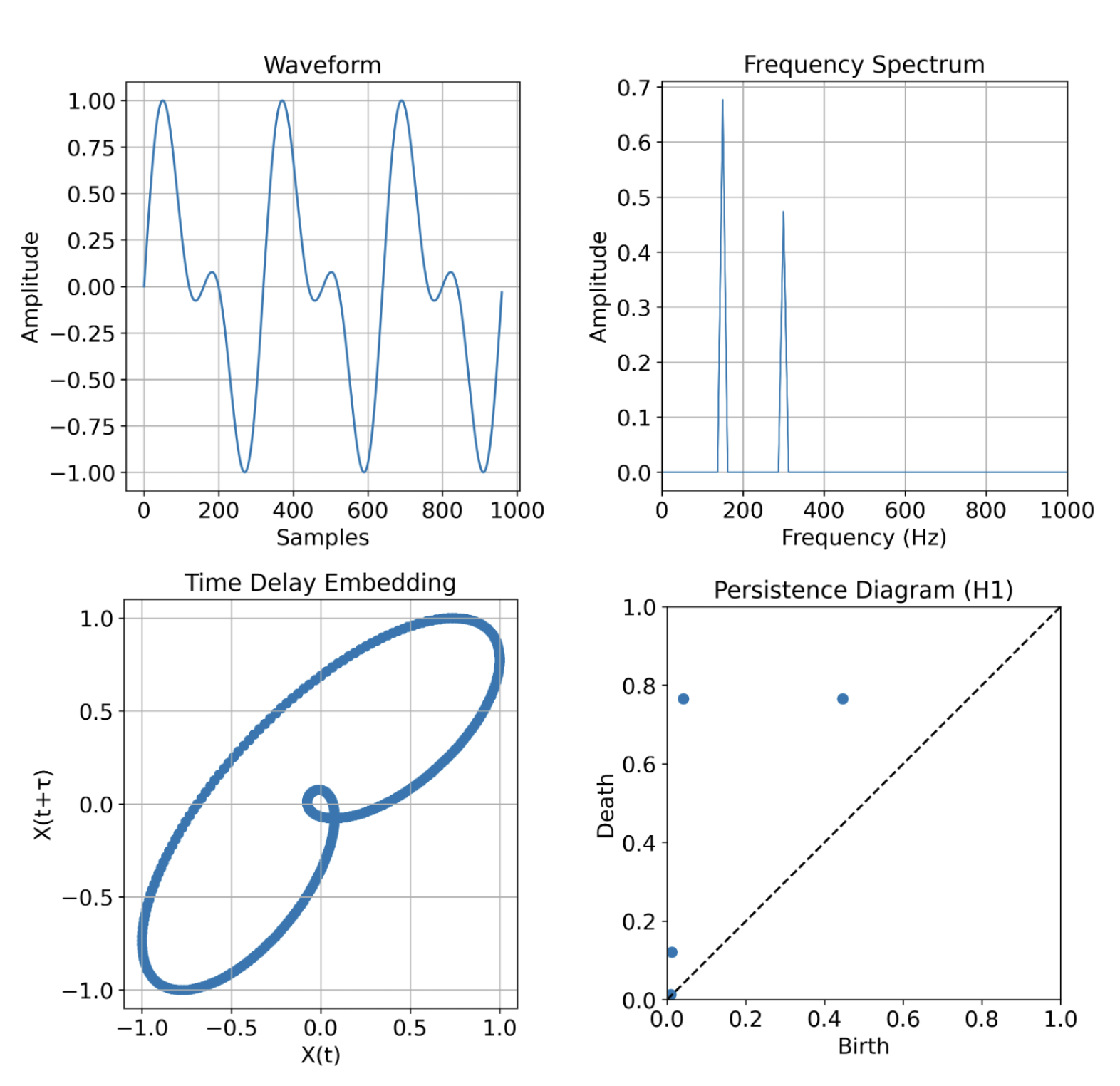}
            \caption{\textbf{TDA of a harmonic signal $s(t)=\sin(2\pi f_0 t) + 0.7\sin(2\pi (2f_0)t)$.} The upper-left panel shows the waveform, the upper-right panel shows the frequency spectrum, the lower-left panel shows the time delay embedding, and the lower-right panel shows the persistence diagram. The addition of the harmonic component introduces a new circular structure in the embedding, which is captured by the persistence diagram as an additional prominent feature.}
    \label{fig:s1_2}
\end{figure}

\begin{figure} [h!] 
    \centering
    \includegraphics[scale=0.5]{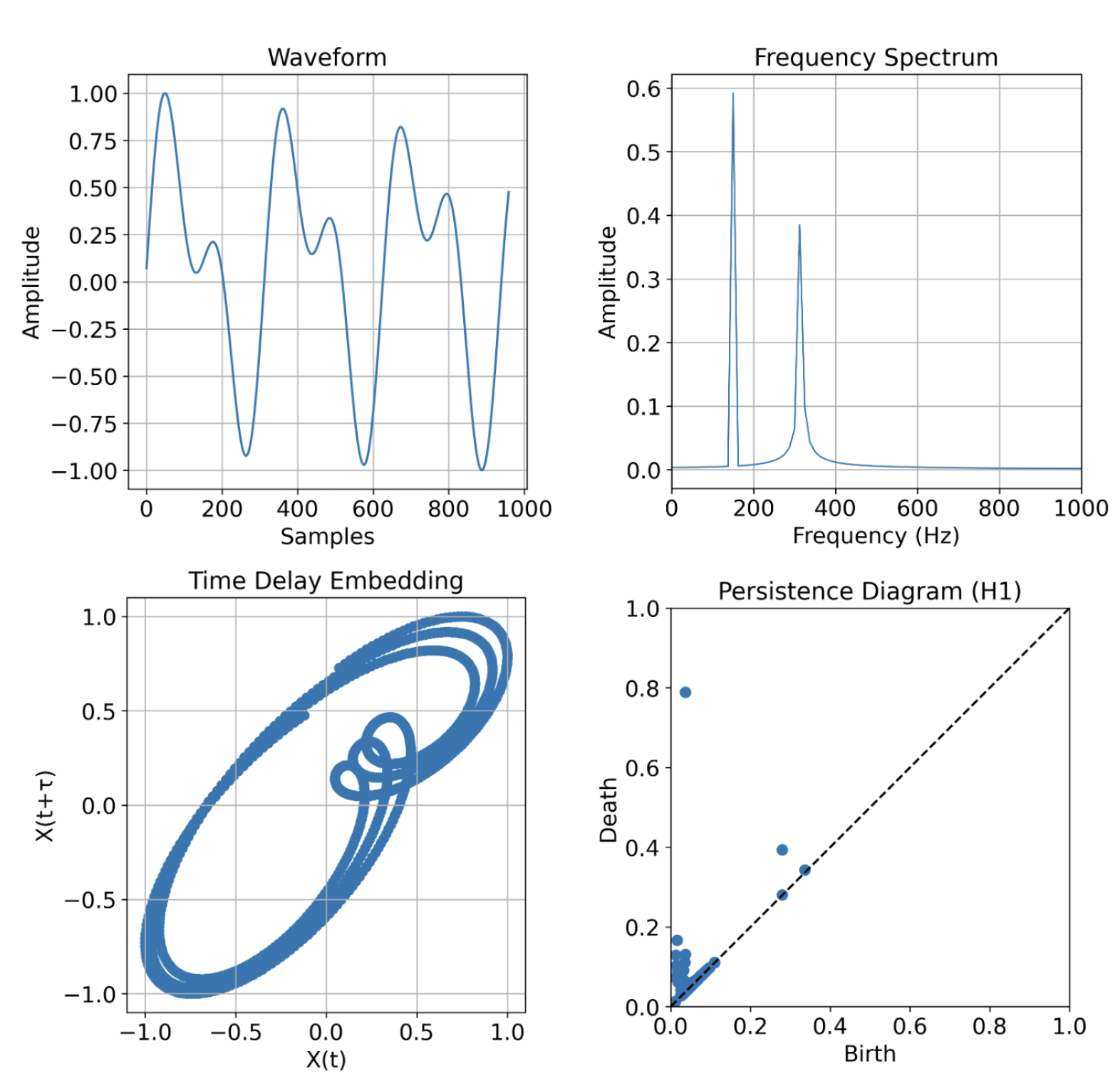}
            \caption{\textbf{TDA of a harmonic signal $s(t)=\sin(2\pi f_0 t) + 0.7\sin(2\pi (2.1f_0)t)$.} The upper-left panel shows the waveform, the upper-right panel shows the frequency spectrum, the lower-left panel shows the time delay embedding, and the lower-right panel shows the persistence diagram. This example includes a harmonic slightly detuned from the integer multiple. Compared with Fig.~\ref{fig:s1_2}, the frequency spectrum shows little difference, but the embedding exhibits a markedly different structure. The persistence diagram captures this change as the emergence of many small circular features.}
    \label{fig:s1_3}
\end{figure}

\begin{figure} [h!] 
    \centering
    \includegraphics[scale=0.5]{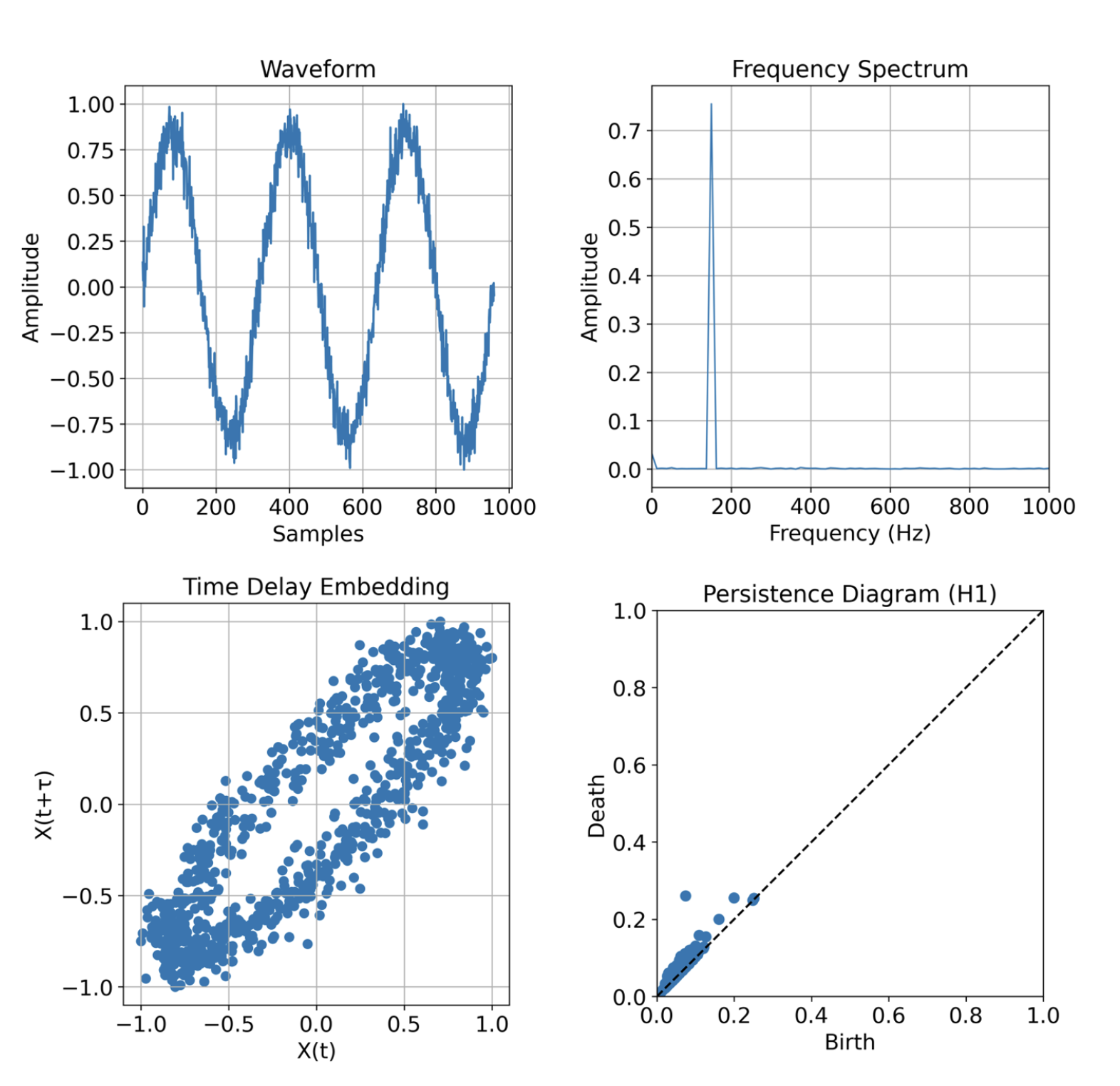}
            \caption{\textbf{TDA of a noisy harmonic signal $s(t)=\sin(2\pi f_0 t) + 0.1\xi(t) \mbox{ with } \xi(t)\sim \mathcal{N}(0,1)$.} The upper-left panel shows the waveform, the upper-right panel shows the frequency spectrum, the lower-left panel shows the time delay embedding, and the lower-right panel shows the persistence diagram. This example adds white noise as a non-integer harmonic component. Consequently, the embedding becomes noisier, and the point representing the circular structure in the persistence diagram moves closer to the diagonal, indicating that the structure is no longer dominant.}
    \label{fig:s1_4}
\end{figure}

We choose the parameters of the input signals as follows: $f_0 = 150\,\mathrm{Hz}$, sampling frequency $48{,}000\,\mathrm{Hz}$, and signal duration $20\,\mathrm{ms}$. The time delay embedding is set to $0.125\,\mathrm{ms}$. The results are shown in Figures~\ref{fig:s1_1}, \ref{fig:s1_2}, \ref{fig:s1_3}, and \ref{fig:s1_4}. In each Figure, the upper-left panel shows the waveform of the analyzed signal, the upper-right panel shows its frequency spectrum obtained via the Fourier transform, the lower-left panel shows the time delay embedding plot, and the lower-right panel presents the persistent diagram.
Figure~\ref{fig:s1_1} illustrates the result for a pure sine wave. In the embedding space, the sine wave forms an elliptical trajectory. TDA applied to this embedding yields a 1-dimensional persistent diagram containing a single prominent point corresponding to the hole structure. Figure~\ref{fig:s1_2} shows the result for a sine wave with an added second harmonic (integer multiple of the fundamental frequency). For waveforms containing integer-order harmonics, the shape of the embedding space changes noticeably. TDA analysis reveals two additional hole structures that emerge during the filtration process. Figure~\ref{fig:s1_3} presents the result when the harmonic frequency is slightly detuned from an exact integer multiple of the fundamental frequency. When the harmonic deviates from the integer ratio, the periodicity in the embedding space is disrupted, and the persistent diagram exhibits numerous small hole structures. These differences are not clearly observable in the frequency spectrum and would be treated as nearly identical by evaluation methods based solely on spectral distribution measures such as sharpness or flatness. This indicates that TDA is effective in capturing such differences in harmonic structure. Finally, Figure~\ref{fig:s1_4} shows the result for a case containing non-integer-order harmonics, simulated by adding white noise to a sine wave. The non-integer harmonics appear as noise in the embedding space and reduce the persistence of the hole structure corresponding to the ellipse observed in the pure sine wave’s persistent diagram. These results demonstrate that, overall, the combination of time delay embedding and TDA can extract harmonic characteristics of audio signals as features in the persistent diagram.

\begin{figure} [h!]
    \centering
    \includegraphics[scale=0.5]{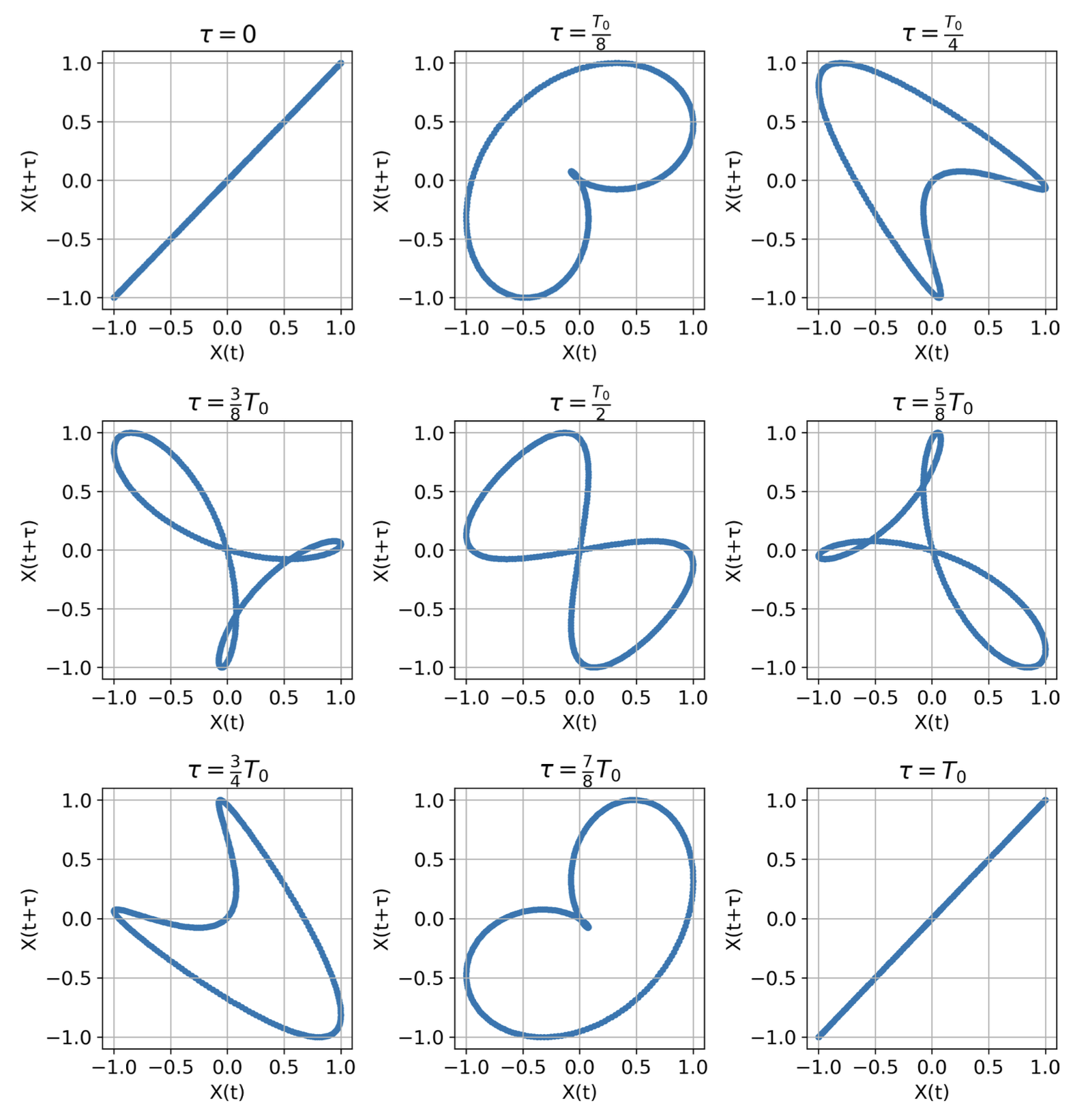}
		\caption{\textbf{Examples of time delay embeddings of the signal $s(t)=\sin(2\pi f_0 t) + 0.7\sin(2\pi (2f_0)t)$ with various time delays.} We can observe that the geometric structure of the embedding space, and, consequently, the topological features, varies significantly depending on the chosen time delay.}
    \label{fig_s2}
\end{figure}

Next, we illustrate how the shape of the embedding space changes when the time delay is varied for a given $s(t)$. Figure~\ref{fig_s2} illustrates the time delay embeddings of the signal $s(t) = \sin(2\pi f_0 t) + 0.7\sin(2\pi(2 f_0)t)$, plotted while varying the time delay $\tau$ from 0 to $2\pi$. As the time delay changes, the number, shape, and overall geometric characteristics of holes in the embedding space vary significantly. This observation indicates that the time delay plays a crucial role in determining the geometric and topological properties captured by the proposed analysis method.

In the following, we investigate the optimization of the time delay based on the proposed method. First, we define artificial signals $s(t; a)$ containing harmonic components as

\begin{equation}
s(t; a) = (1-a) \sin(2\pi f_0 t) + a \left( \sum_{n=1}^N A(n) \sin(2\pi n f_0 t) + \mathcal{F}^{-1} \left( B(f) \cdot \mathcal{F}(\xi(t)) \right) \right),
\label{eq:synthesis}
\end{equation}

\noindent where the first term is a sine wave at the fundamental frequency $f_0$, and the second term contains both harmonic and non-integer-order harmonic components. In the second term, the first part represents the integer-order harmonics, with $A(n)$ denoting the amplitude of the $n$-th harmonic. The second part represents the non-integer-order harmonics, where $B(f)$ specifies the frequency-dependent noise weight. Here, $\xi(t)$ is white noise, and $\mathcal{F}$ and $\mathcal{F}^{-1}$ denote the Fourier transform and its inverse, respectively. For $A(n)$ and $B(f)$, we adopt the following seven patterns commonly used in audio synthesis and signal processing:

\begin{itemize}
    \item \textbf{Triangle wave}: $A(n) = \frac{1}{n^2}$, \quad $n = 1, 3, 5, \dots$, \quad $B(f) = 0$;
    \item \textbf{Square wave}: $A(n) = \frac{1}{n}$, \quad $n = 1, 3, 5, \dots$, \quad $B(f) = 0$;
    \item \textbf{Sawtooth wave}: $A(n) = \frac{1}{n}$, \quad $n = 1, 2, 3, \dots$, \quad $B(f) = 0$;
    \item \textbf{Modified sawtooth wave}: $A(n) = \frac{1}{n^2}$, \quad $n = 1, 2, 3, \dots$, \quad $B(f) = 0$;
    \item \textbf{White noise}: $A(n) = 0$, \quad $B(f) = 1$;
    \item \textbf{Pink noise}: $A(n) = 0$, \quad $B(f) = \frac{1}{\sqrt{f}}$;
    \item \textbf{Brown noise}: $A(n) = 0$, \quad $B(f) = \frac{1}{f}$.
\end{itemize}

\begin{figure} [h!]
     \centering
     \includegraphics[scale=1.5]{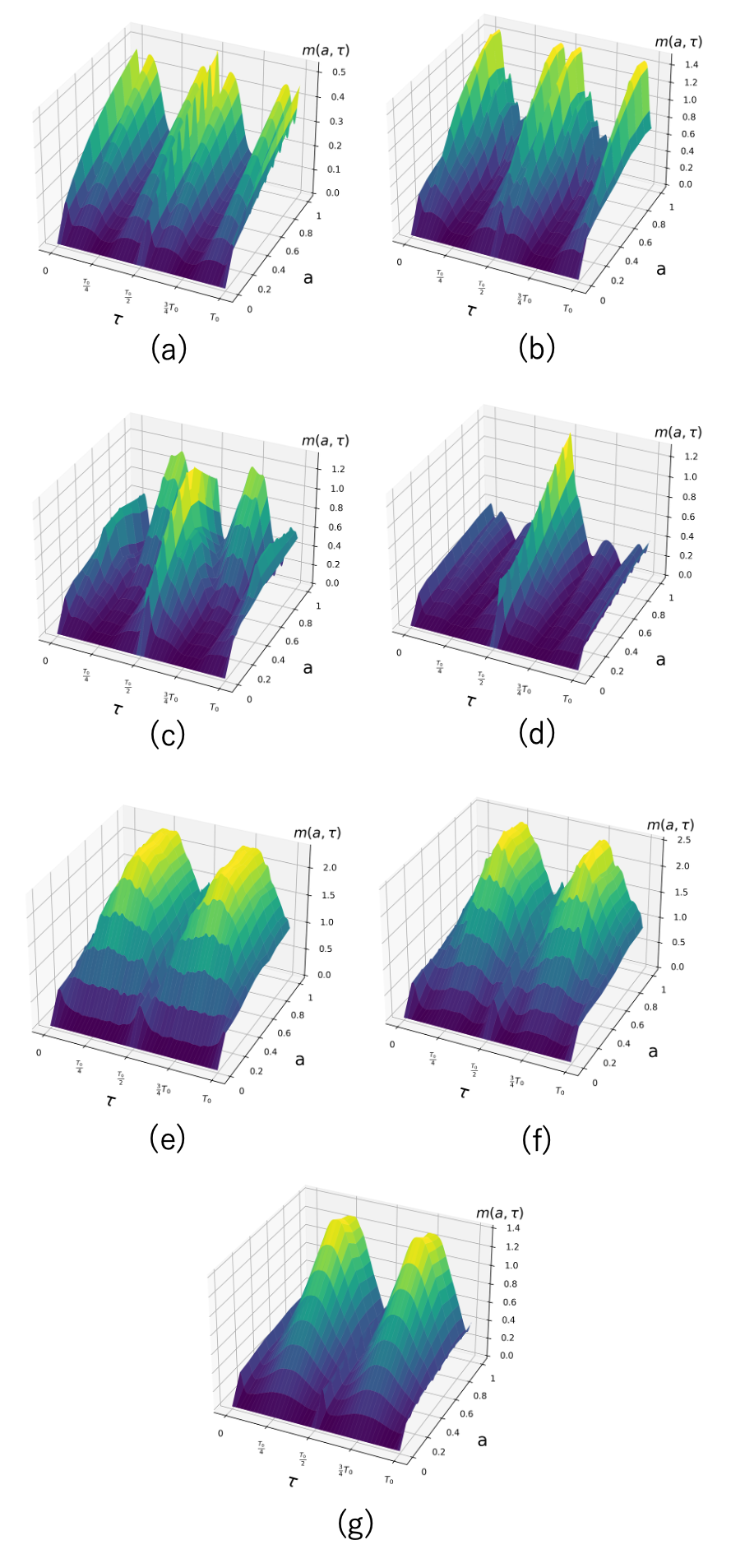}
     \caption{\textbf{Variation of the topological feature for different synthesized waveform.} Results for the following waveforms: (a) Triangle wave, (b) Square wave, (c) Sawtooth wave, (d) Modified sawtooth wave, (e) White noise, (f) Pink noise, and (g) Brown noise.
The x-axis represents the time delay, the y-axis indicates the harmonic strength, and the z-axis shows the magnitude of the topological feature.
By varying the time delay, it can be observed that the rate of increase in the topological feature with respect to the harmonic strength differs among waveforms, reflecting the distinct types of harmonics inherent to each waveform.}
     \label{fig_s3}
\end{figure}

For each of these waveforms, we compute the topological feature $m$ while varying both the overtone strength $a$ and the time delay embedding $\tau$. The results are shown in Fig.~\ref{fig_s3}, where the relationship between $m$, $a$, and $\tau$ is visualized. The Figure illustrates the variation of the topological feature for each synthesized waveform: (a) Triangle wave, (b) Square wave, (c) Sawtooth wave, (d) Modified sawtooth wave, (e) White noise, (f) Pink noise, and (g) Brown noise. The $x$-axis represents the time delay embedding $\tau$, the $y$-axis indicates the overtone strength $a$, and the $z$-axis shows the magnitude of the topological feature $m$. The time delay was discretized in $1\,\mathrm{Hz}$ steps, and the overtone strength was discretized in $10\%$ increments. For each set of parameters, the topological feature was computed according to Eq.~\ref{eq:pm2}. The sampling frequency was set to $48{,}000\,\mathrm{Hz}$, and one period of $s(t)$ was used for analysis. The harmonic components were limited to a maximum of the 10th order. Experiments were conducted with three different fundamental frequencies, $f_0 = 150\,\mathrm{Hz}$, $250\,\mathrm{Hz}$, and $500\,\mathrm{Hz}$. Since similar trends were observed for all $f_0$, only the results for $f_0 = 150\,\mathrm{Hz}$ are presented. The results show that, for each synthesized signal, the topological feature $m(a,\tau)$ tended to increase with increasing overtone strength $a$. This confirms that the proposed analysis method can effectively capture changes in the harmonic structure. Furthermore, the growth pattern of $m$ with respect to $a$ was found to vary considerably depending on the choice of $\tau$. However, no single optimal time delay $\tau$ common to all signals was identified.  In particular, two distinct patterns were observed:  (i) signals for which the change in $m$ was maximized when $\tau = T_0/2$ (where $T_0$ is the fundamental period), and  (ii) signals for which the change in $m$ was maximized when $\tau = 3T_0/4$ or $\tau = 4T_0$.

 \begin{figure} [h!]
     \centering
     \includegraphics[scale=0.5]{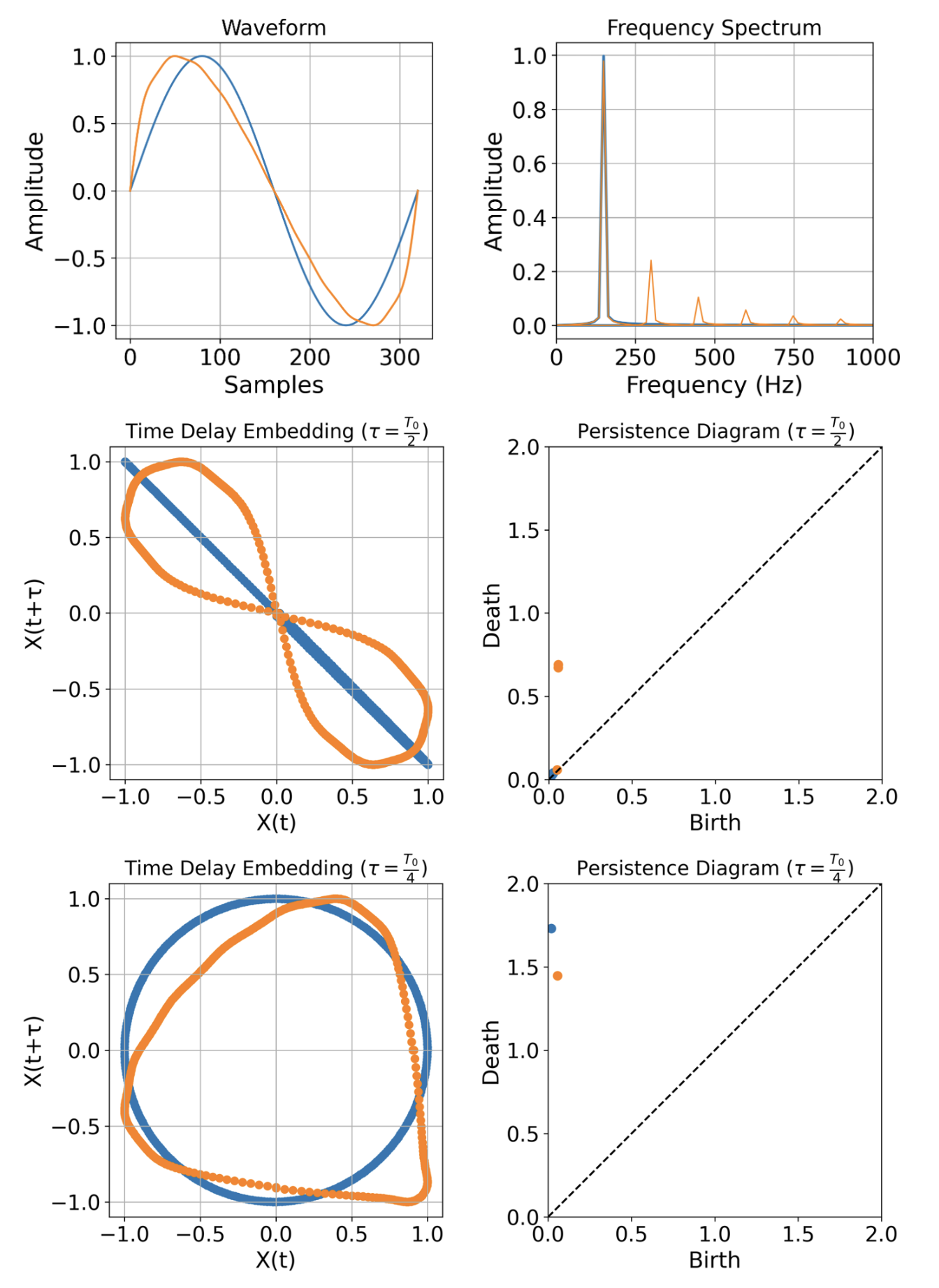}
     \caption{\textbf{Embedding space and corresponding persistent diagram for a modified sawtooth wave.} The blue line represents a sine wave without harmonics, and the orange line represents a modified sawtooth wave containing integer harmonics. The upper-left panel shows the waveform, the upper-right panel shows the frequency spectrum, the middle-left and middle-right panels show the time delay embedding and the corresponding persistence diagram for $\tau = T_0 / 2$, respectively, and the lower-left and lower-right panels show those for $\tau = T_0 / 4$.
The topological feature magnitudes are $m(1, T_0 / 2) = 1.29$ and $m(1, T_0 / 4) = 0.29$, indicating that the addition of harmonics produces a more pronounced qualitative change in the geometric structure of the embedding space when $\tau = T_0 / 2$.}
     \label{fig_s4}
 \end{figure}

As an example, Fig.~\ref{fig_s4} illustrates the time delay embedding space and the corresponding persistence diagram for a modified sawtooth wave. In the Figure, blue markers represent pure sine waves, while orange markers correspond to waveforms containing harmonics. The upper-left panel shows the input waveform, the upper-right panel presents the frequency spectrum obtained via the Fourier transform, the center-left panel shows the embedding for $\tau = T_0/2$, and the center-right panel presents its corresponding persistent diagram. The lower-left panel shows the embedding for $\tau = T_0/4$, and the lower-right panel shows its persistent diagram.

As previously shown, when a sine wave is delayed by $T_0/2$, the symmetry of the waveform results in a straight-line trajectory in the embedding space. Then, when harmonics are added, this symmetry is broken, and hole structures emerge even for the same time delay.  In contrast, for $\tau = T_0/4$, the overall embedding shape changes only gradually, and no significant topological changes are observed. This tendency was observed for waveforms containing integer-order harmonics. Therefore, integer-order harmonics are considered to modify the topological features by altering the geometric structure of the embedding space and introducing hole structures that do not exist in pure sine waves.
This effect is particularly pronounced when $\tau = T_0/2$. It is also noteworthy that, as shown in Fig.~\ref{fig_s3}(b), for certain waveforms the variation in the topological feature is maximized not exactly at $\tau = T_0/2$, but at a time delay slightly offset from $T_0/2$.  This can be explained by the fact that, even when the waveform is significantly altered by harmonics, if the time delay embedding is exactly $T_0/2$, any component retaining the same $T_0/2$ symmetry as a sine wave will still be plotted as a straight line.  However, for such waveforms, the largest topological changes—namely, the addition of new hole structures—still occur in the vicinity of $\tau = T_0/2$, where the original sine wave embedding does not exhibit large hole structures. From a practical perspective, it is therefore preferable to select a time delay slightly offset from $T_0/2$.

\begin{figure} [h!]
    \centering
    \includegraphics[scale=0.5]{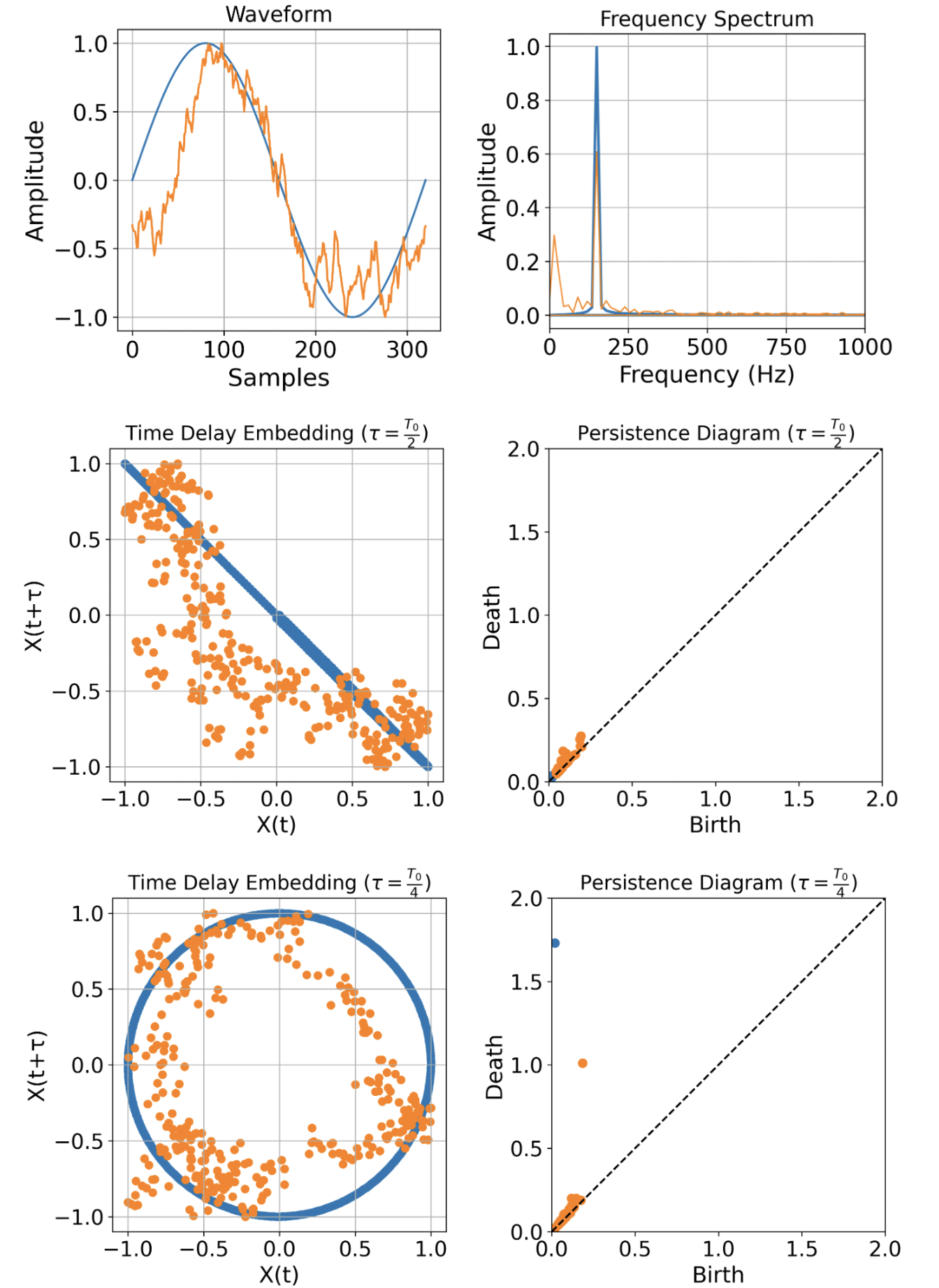}
    \caption{\textbf{Embedding space and corresponding persistent diagram for a waveform from Brown noise.} The blue line represents a sine wave without harmonics, and the orange line represents Brown noise containing non-integer harmonics. The upper-left panel shows the waveform, the upper-right panel shows the frequency spectrum, the middle-left and middle-right panels show the time delay embedding and the corresponding persistence diagram for $\tau = T_0 / 2$, respectively, and the lower-left and lower-right panels show those for $\tau = T_0 / 4$.
The topological feature magnitudes are $m(0.5, T_0 / 2) = 1.08$ and $m(0.5, T_0 / 4) = 1.69$, indicating that although non-integer harmonics introduce noise in both cases, $\tau = T_0 / 2$ shows little change in geometric structure since no circular structure originally existed, whereas $\tau = T_0 / 4$ exhibits a reduction in the dominance of the circular structure that was originally present.}
    \label{fig_s5}
\end{figure}

Fig.~\ref{fig_s5} shows the embedding space and the corresponding persistent diagram for a waveform generated using Brown noise. When non-integer-order harmonics such as Brown noise are added, noise is introduced into the embedding trajectory, resulting in a reduction of the persistence of the corresponding features in the persistent diagram.  
For time delays of $\tau = 3T_0/4$ and $\tau = 4T_0$, the sine wave is plotted as a circular trajectory in the embedding space.  
This embedding configuration yields the largest persistence of hole structures among all time delays.  
Therefore, since this embedding best captures the reduction in persistence caused by the added noise, it can be considered the optimal time delay for analyzing waveforms containing non-integer-order harmonics. In contrast, when $\tau = T_0/2$, the sine wave does not possess a circular structure to begin with; therefore, the shape deformation caused by the non-integer-order harmonics does not significantly affect its topological features.

From these results, it can be concluded that in harmonic analysis using TDA, the embedding configuration has a significant impact on the outcome, and that performing time delay embeddings with different time delays allows one to focus on different harmonic characteristics.

\subsection{Real data}
In this Section, we validate the appropriateness of the optimal time delays obtained using the proposed method through experiments on real-world audio data. For the dataset, we use the NSynth dataset~\cite{nsynth2017}, which consists of four-second audio recordings of one-shot notes from 1,006 different instruments, sampled at $16{,}000\,\mathrm{Hz}$. Each instrument is categorized into one of the following groups: bass, brass, flute, guitar, keyboard, mallet, organ, reed, string, synth lead, and vocal.

Based on the results of the experiments with synthesized data, we compute the topological feature values using the proposed method with time delays $\tau = T_0/2$ and $\tau = T_0/4$. The fundamental frequency for analysis is set to C4 in the MIDI scale ($f_0 = 261.6\,\mathrm{Hz}$). When applying the proposed method to real data, it is necessary to select an analysis segment of appropriate length.  If the segment is too long, the number of points in the embedding space increases, which significantly increases the computation time of TDA.  Conversely, if the segment is too short, the harmonic characteristics may not be accurately captured.  Based on preliminary experiments, we extract a segment corresponding to four periods of the fundamental frequency for analysis.  Furthermore, the start time of the segment is set to the moment when the amplitude reaches its maximum, which is assumed to correspond to the point where the harmonic structure of the instrument is most prominent.

 \begin{figure}[h!]
    \centering
    \includegraphics[scale=0.5]{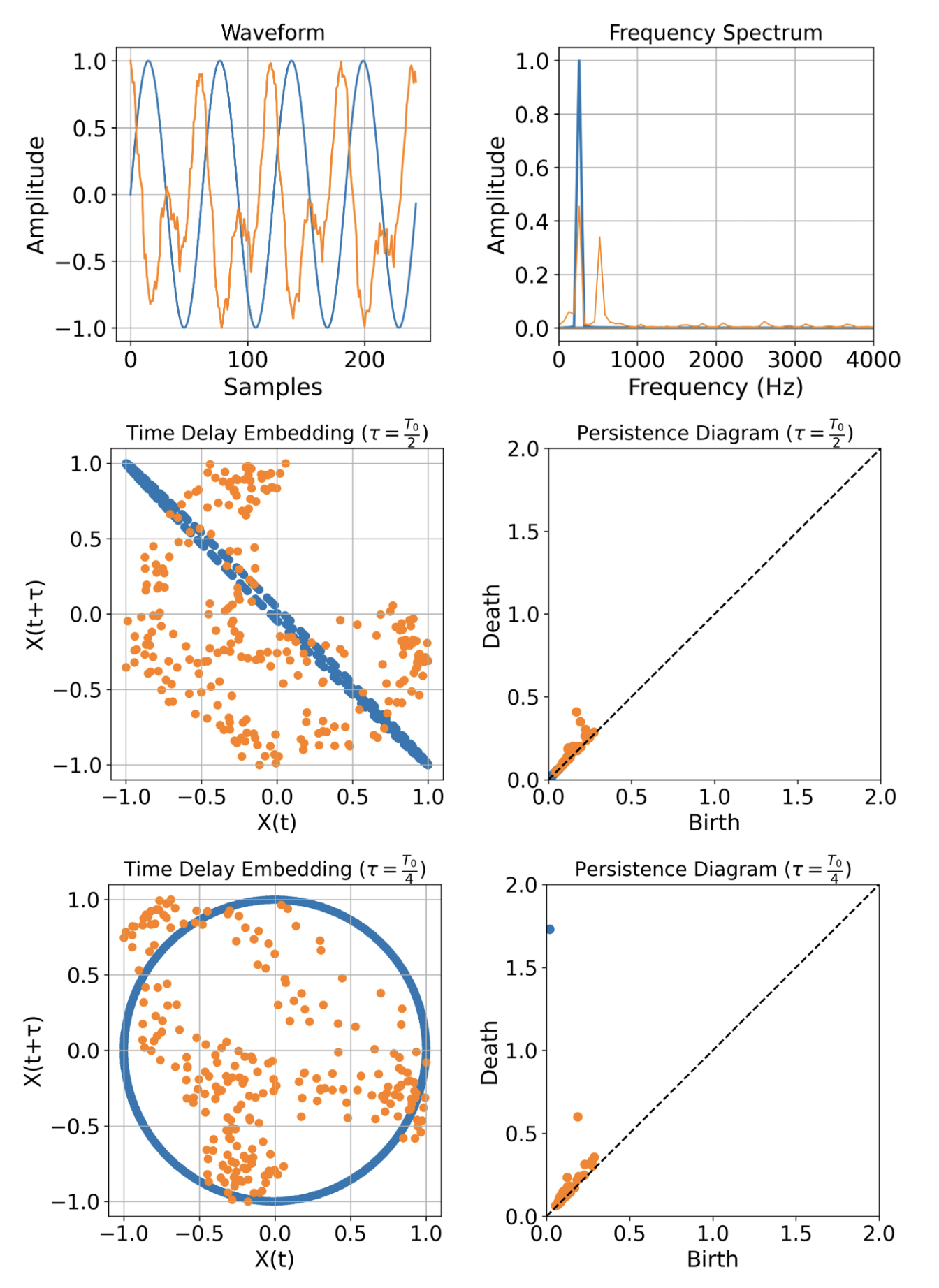}
    \caption{\textbf{Embedding space and corresponding persistent diagram for a waveform from a guitar.} The blue line represents a sine wave without harmonics, and the orange line represents the analysis result of a guitar sound from the dataset. The upper-left panel shows the waveform, the upper-right panel shows the frequency spectrum, the middle-left and middle-right panels show the time delay embedding and the corresponding persistence diagram for $\tau = T_0 / 2$, respectively, and the lower-left and lower-right panels show those for $\tau = T_0 / 4$.
The topological feature magnitudes are $m(T_0 / 2) = 0.92$ and $m(T_0 / 4) = 2.00$, indicating that this sound source contains relatively dominant non-integer harmonics.}
    
    \label{fig_r1}
\end{figure}

As an example, we show the analysis results for a guitar waveform using the proposed method, depicted in Fig.~\ref{fig_r1}. There, the blue plots correspond to sine waves, while the orange plots correspond to waveforms containing harmonics. The upper-left panel shows the input waveform, the upper-right panel presents the frequency spectrum obtained via the Fourier transform, the center-left panel shows the embedding for $\tau = T_0/2$, and the center-right panel presents its corresponding persistent diagram. The lower-left panel shows the embedding for $\tau = T_0/4$, and the lower-right panel presents its persistent diagram.
These results demonstrate that the proposed method can also extract timbral features from real-world data.  For the guitar sound, Although the topological feature values are relatively large for both time delays, they are particularly large for $\tau = T_0/4$.  This suggests that the analyzed signal contains a substantial amount of non-integer-order harmonics.

 \begin{figure} [h!]
     \centering
     \includegraphics[scale=0.5]{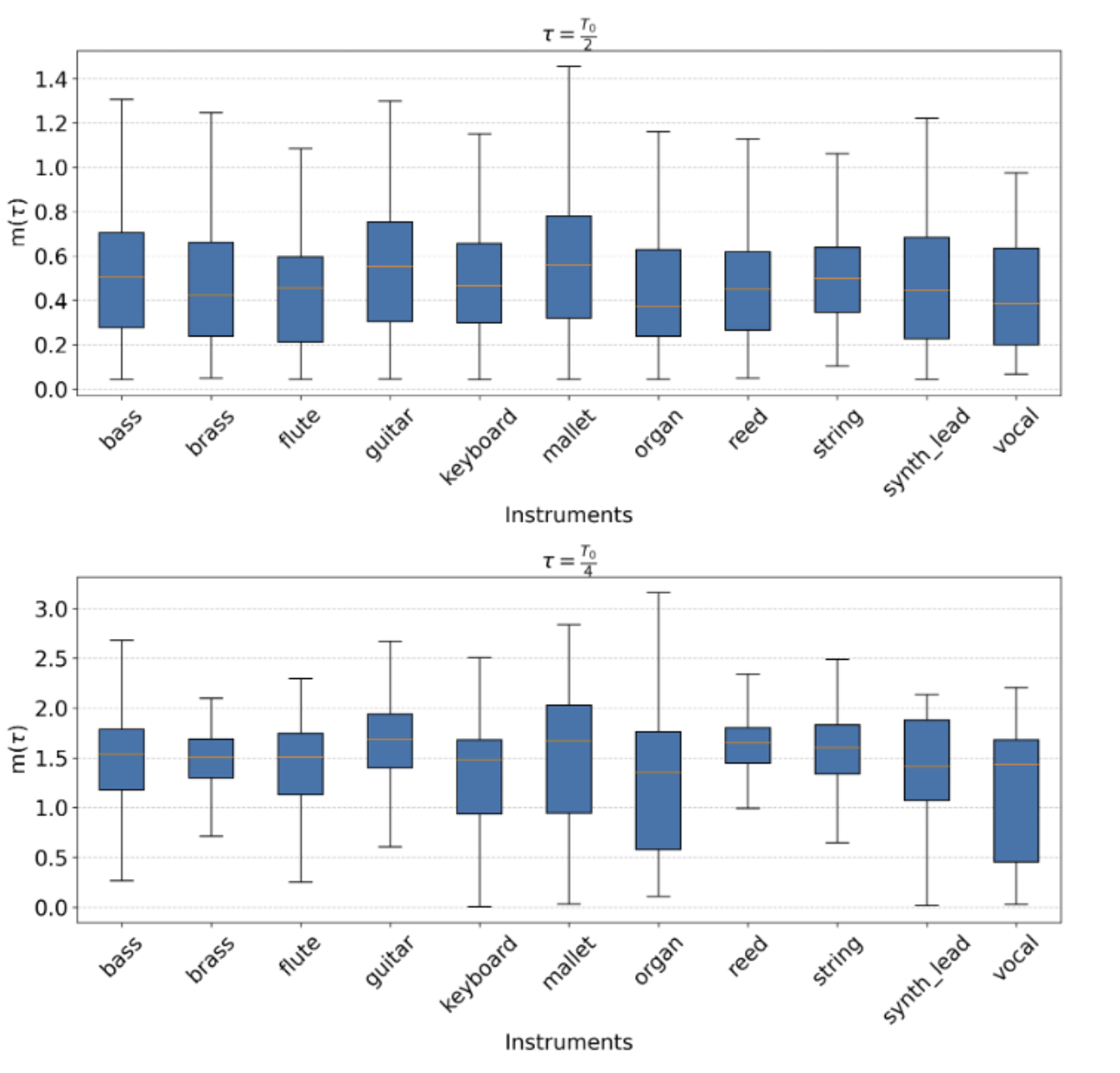}
     \caption{\textbf{Distributions of the topological feature values across different instrument categories for two different time delay embeddings.} The proposed method is applied to each sound source in the dataset, and the distribution of the resulting topological feature magnitudes is shown. The x-axis represents the instrument categories defined in the dataset metadata, and the y-axis represents the values of the topological feature. The upper plot shows the results embedded with $\tau = T_0 / 2$, focusing on integer harmonics, while the lower plot shows those embedded with $\tau = T_0 / 4$, focusing on non-integer harmonics. The results demonstrate that the non-integer harmonic representation captures inter-instrument differences more clearly.}
     \label{fig_r2}
 \end{figure}

Next, we apply the proposed method to all instrument samples in the dataset and compare the distributions of the topological feature values across instrument categories. The results of this analysis are shown in Fig.~\ref{fig_r2}. There, the x-axis represents the instrument categories defined in the dataset metadata, and the y-axis represents the values of the topological feature. The upper plot shows the results embedded with $\tau = T_0/2$, focusing on integer harmonics, while the lower plot shows those embedded with $\tau = T_0/4$, focusing on non-integer harmonics. It can be observed that each instrument category exhibits a distinct distribution of topological feature values.  
For instance, the flute shows small topological feature values for $\tau = T_0/2$, indicating a low presence of integer-order harmonics, while the guitar exhibits large values for both time delays, suggesting that it contains a substantial amount of both integer- and non-integer-order harmonics.  These observations are consistent with the known harmonic characteristics of the respective instruments. On the other hand, there are a number of instrument categories for which no significant differences are observed in the present comparison.  For these instruments, the differences in timbre are likely influenced not only by the local harmonic characteristics but also by their temporal variations.  
Therefore, as a future extension of this work, the proposed method could be applied repeatedly to multiple segments of the same signal to track the temporal evolution of the topological features of the harmonics.

\section{Conclusions}\label{sec:conc&persp}

In this work, we have introduced a method for selecting suitable time delays to analyze harmonic structures using time delay embedding together with Topological Data Analysis (TDA). Through numerical experiments with artificial signals, we have demonstrated that time delay embedding combined with TDA can successfully capture harmonic characteristics. We have also observed that the geometry of the embedded space strongly depends on the chosen time delay. By examining topological features, we have found that the time delays $T_0/2$ and $T_0/4$, where $T_0$ is the fundamental period, are particularly useful. Specifically, using $T_0/2$ emphasizes the fundamental harmonic, while $T_0/4$ highlights the presence of both integer and non-integer harmonics. Additional experiments with real audio data have shown that choosing time delays according to our proposed method can also be effective for analyzing harmonic structures in real musical instrument sounds, opening the way for several applications in both fundamental research and industry.

There are several directions for future works. One important task is to adapt the method more effectively to real-world data, for example by developing ways to follow changes in timbre over time and by reducing the computational cost of the analysis. In this study, we focused on topological features that measure deviations from a pure sine wave. A more comprehensive evaluation might be achieved by incorporating additional statistical information from the persistence diagrams, such as average lifetime or variance of features, which remains an open problem. Another promising direction is to optimize embeddings in higher dimensions, which we have not explored here. A recent work in hierarchical embedding methods~\cite{zhang2024hierarchical} suggests that this could enable the analysis of more complex sounds, including chords and rich tonal structures, and that could be applied to the framework hereby developed. \\

\paragraph*{Acknowledgements}
R.M. acknowledges Stefanella Boatto, Conrado Catarcione, Maria Gabriella Cavalcante Basilio, and Victor Pessanha for discussions on topology and TDA.\\

\paragraph*{Data availability}
The interested reader may contact G.S. at gssato.ac@gmail.com for further information. The code to make this study is available at \url{https://github.com/gssato/TopologicalHarmonicExtraction}. \\

\paragraph*{Funding}
G.S. and H.N. acknowledge H.N. acknowledges JSPS KAKENHI 25H01468, 25K03081, 22H00516 and 22K11919 for financial support. R.M. acknowledges JSPS, grant 24KF0211.\\

\paragraph*{Conflict of interest}
The authors declare no competing interests.\\

\paragraph*{Author contributions}
\begin{tabular}{ll}
Conceptualization: & G.S., R.M.\\
Investigation: & G.S. \\
Visualization: & G.S., R.M. \\
Software: & G.S.\\
Supervision: & H.N., R.M.\\
Project administration: & R.M. \\
Funding acquisition: & H.N.\\
Writing--original draft: & G.S., R.M.\\
Writing--review \& editing: & G.S., H.N., R.M.
\end{tabular}

\end{document}